# HIGGS BOSON – ON YOUR OWN

T. Csörgő

MTA Wigner Research Center for Physics, Institute for Particle and Nuclear Physics

H- 1525 Budapest 114, POBox 49, Hungary

**Abstract:**

One of the highlights of 2012 in physics is related to two papers, published by the ATLAS and the CMS Collaborations at the CERN LHC. These two experiments announced the discovery of at least one new particle in p+p collisions at CERN LHC. At least one of the properties of this new particle is found to be similar to that of the so called Higgs boson, the last and most elusive particle from the Standard Model of particle physics. Physics teachers are frequently approached by their media-educated students, who inquire about the properties of the Higgs boson, but physics teachers are rarely trained to teach this elusive aspect of particle physics in elementary, middle or junior high schools.

In this paper I describe a card-game, that can be considered as a hands-on and easily accessible tool that allows interested teachers, students and also motivated lay-persons to play with the properties of the newly found particle. This new particle was experimentally found through its decays to directly observable, final state particles. Many of these final state particles are represented in a deck of cards, that represent elementary particles, originally invented to popularize the physics of quark matter in Quark Matter Card Games. The Higgs decay properties can be utilized, playfully, in a Higgs boson search card game. The rules of this game illustrate also the needs for a realistic element of some luck, to complement knowledge and memory, useful skills that this game also helps to develop. The paper is organized as a handout or booklet, that directly describes, how to play the "Higgs boson – on Your Own" card game.

# HIGGS BOSON – ON YOUR OWN

**Number of players:** in principle, arbitrary, but in practice, limited by available space at a table.

**Object of the game:** to win, by detecting a decay of a Higgs boson. If this does not happen in a given game, one can win by statistics, by collecting the largest number of particle cards.

**The course of the game:** The players put the thoroughly shuffled pack to the middle then make an ordered, e.g. 6x11 table of the cards, that all are placed in a face down position, initially. Players follow each other in a sequential, e.g. counter-clockwise order. This Higgs-boson – on Your Own is a Memory style game, it starts similarly to the Memory of Quark Matter Card Game.

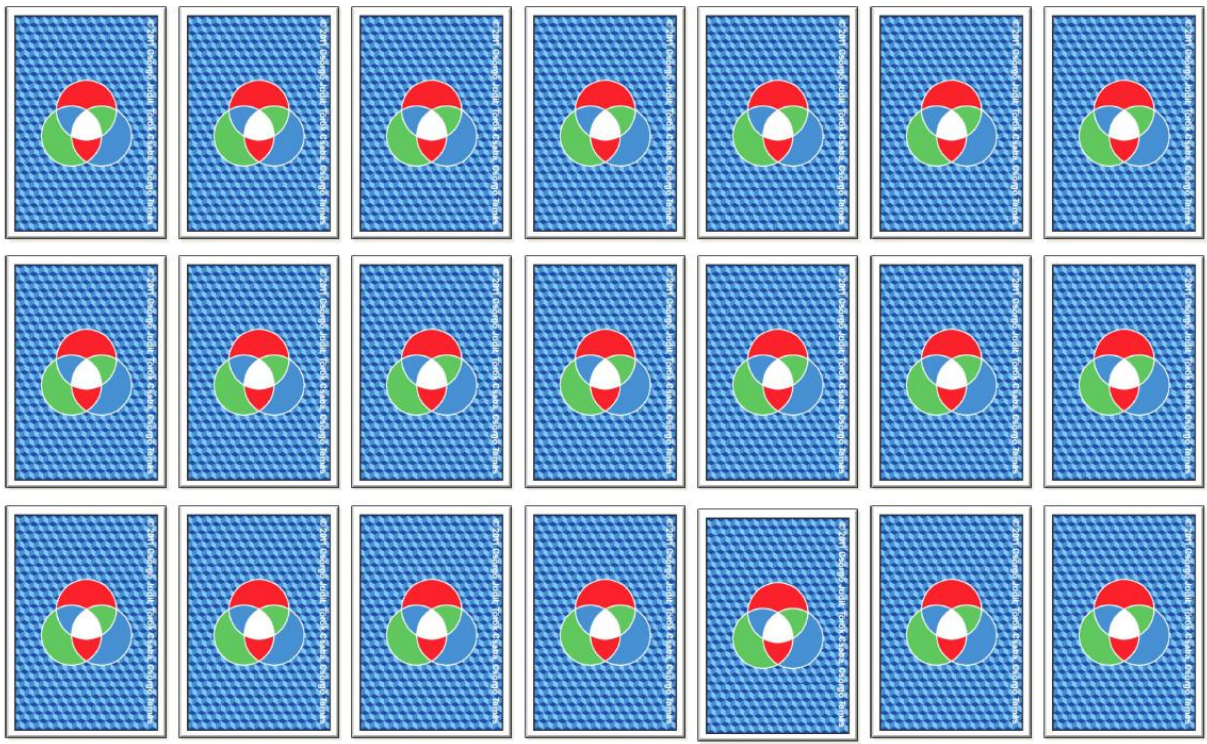

**Figure 1.** Illustration for an initial configuration for the Higgs Boson – on Your Own card game. As the deck has 66 cards, that represent colored (anti)quarks as well as black and white leptons, a realistic situation starts with a 6 x 11 table of cards, all in a face down position.

Players may form lepton-antilepton pairs, various mesons and baryons from the cards, and may also detect a decay of a Higgs boson. The winner is that player, who is the first to find a valid Higgs boson, via its decay, or, if Higgs decay is not realized in a particular course of the game, the winner is the one, who collected most of the cards. The first player turns two cards face up. In contrast to the usual "Memory" style games, where players can keep the cards if both are the same, in the Higgs boson – on Your Own, or in a Quark Matter Memory games, the players can keep cards only if a valid particle-antiparticle pair (or, if a valid set of three or four cards) is formed. Valid pairs come in two forms: either two black-and-white cards that represent lepton pairs, namely electron-positron, muon-antimuon or neutrino-anti-neutrino pairs, or if they can form hadrons from two or three colored quark and anti-quark cards. Finding an electron (e-) and positron (e+) pair is illustrated on Figure 2. If a valid pair of particles is found, the player removes them from the table and keeps them, by placing these cards before him- or herself.

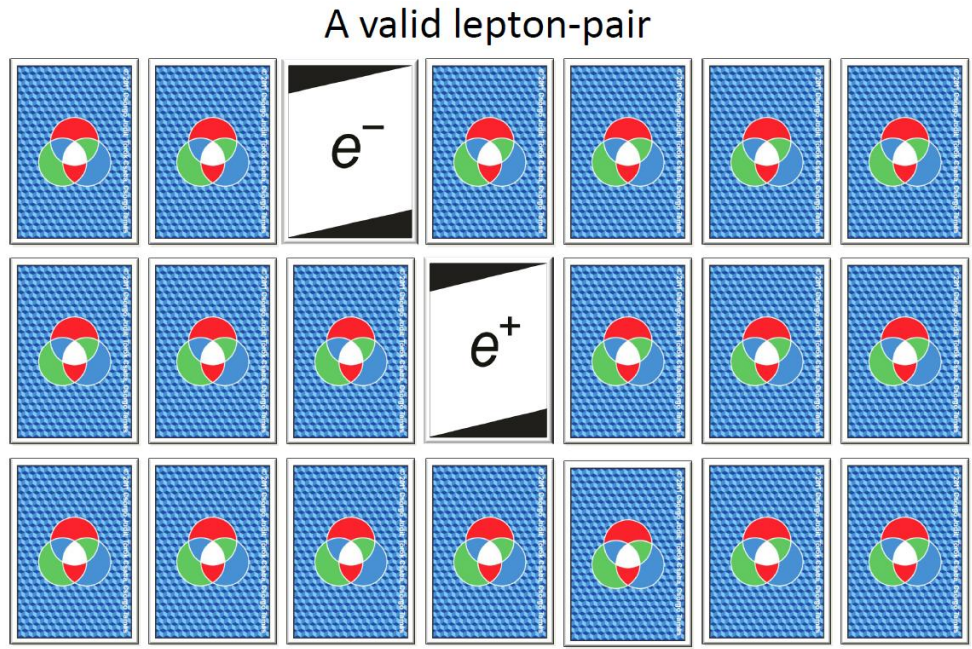

**Figure 2.** Illustration of finding a valid pair of lepton-antilepton cards in the Higgs Boson – on Your Own game. These cards are removed and the player may try to find another valid combination of cards again – and in case of success – again. This series lasts until the cards that are turned face up do not anymore correspond to a valid combination. Then the next player starts to look for valid card combinations, and continues till failure, and so on and so on.

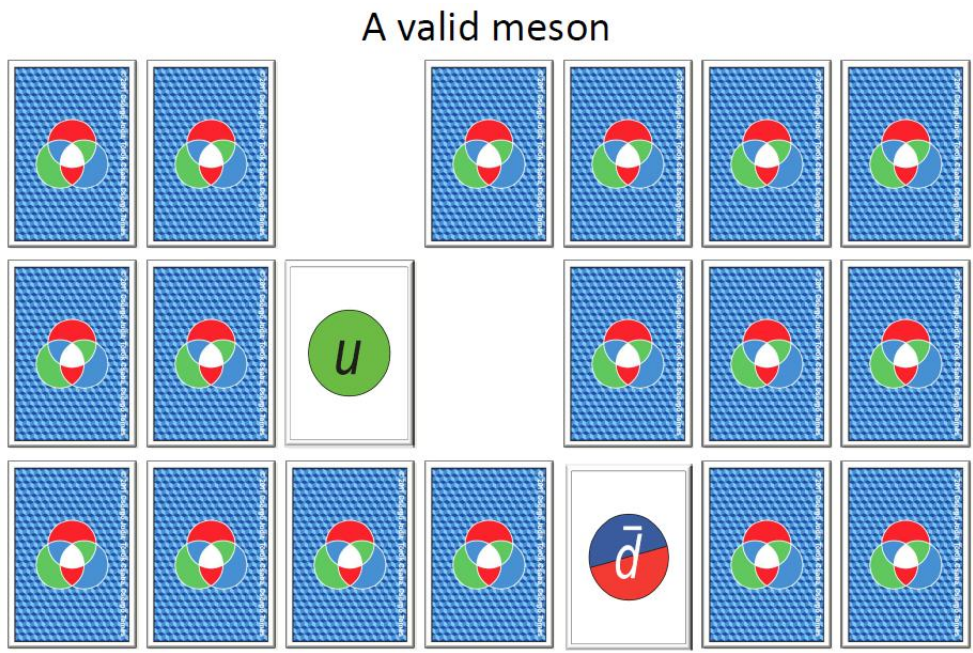

**Figure 3.** Finding a meson (a pair of quark – antiquark cards, where all the three of red, green and blue colors are represented) in the Higgs Boson – on Your Own card game.

Hadrons may come in two forms: mesons and (anti)baryons. Both can be formed from the colored (anti)quark cards, modeling the formation of hadrons, which are the observable and strongly interacting particles. Mesons are formed from quark-antiquark cards. Baryons are formed from three quark cards. The quark cards have either a red, a green or a blue color, while the antiquarks are marked with two colors at the same time, as a red/green, green/blue or blue/red combinations. Valid mesons and baryons too feature all the three (red, green, blue) colors. Figure 3 illustrates a valid pair of colored cards, that corresponds to the finding a meson. A valid combination of three quarks (red, green and blue quarks) corresponds to the finding a baryon, as shown in Figure 4.

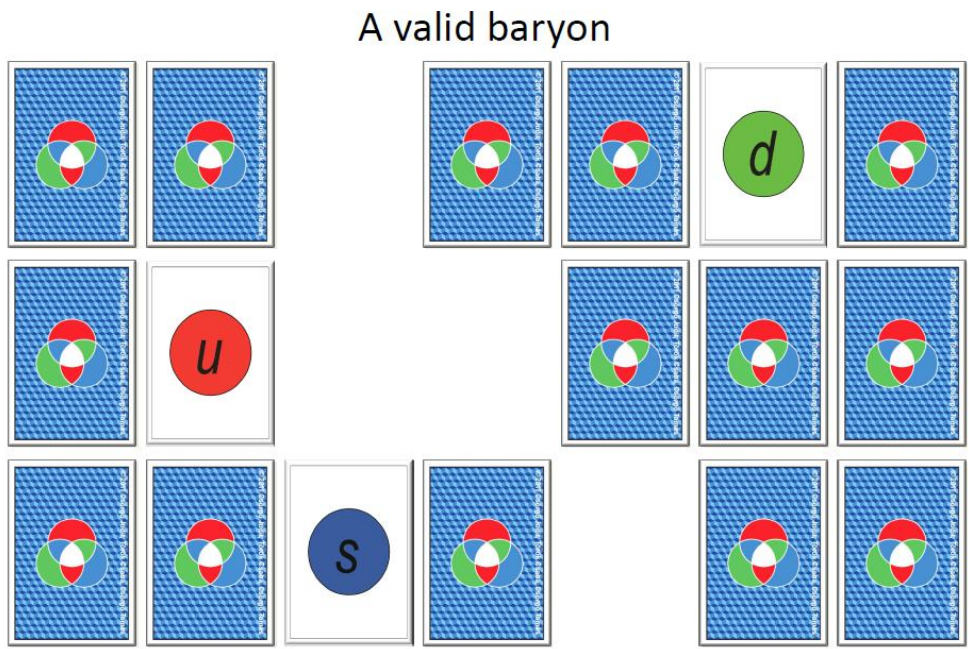

**Figure 4.** Finding a baryon (a combination of three colored quark cards, with each of a red, green and blue color) in the Higgs Boson – on Your Own card game. After turning up two quarks with different color (e.g. a red u and a green d) the player realizes that this can be completed to a baryon, so he can turn up another card (e.g. a blue s quark). In case of success, like the one shown in this illustration, the cards are removed from the table and the player may continue to turn cards to face up position. If in a given round a valid combination is not found, the player has to put back the cards to their original place, in a face down position, and the next player may try to find a valid combination of cards.

If the first two cards are both (anti)quarks (both represented with a red, a green or a blue colored card of if both are two-colored) and if their colors are different, the players may turn a third card to a face up position, to try to form a baryon (or, an anti-baryon). Players can keep all the three cards if they are all quarks (or all anti-quarks) and all the three colors are represented equally. If the pairs or triplets of cards are not valid combinations, than players have to return them to their original place in the table, again to a face down position. Players follow each other in an agreed (e.g. clock-wise) order untill all the cards are collected. The goal is to collect the largest number of cards, or, to find a Higgs boson and to win the game this way. Without the Higgs boson search, the game described until now is the so called Memory style Quark Matter Card game. At the end of the game, a small mathematical theorem quarantees that all the cards disappear from the table if none of the players was cheating. The winner is that player, who collected most of valid cards.

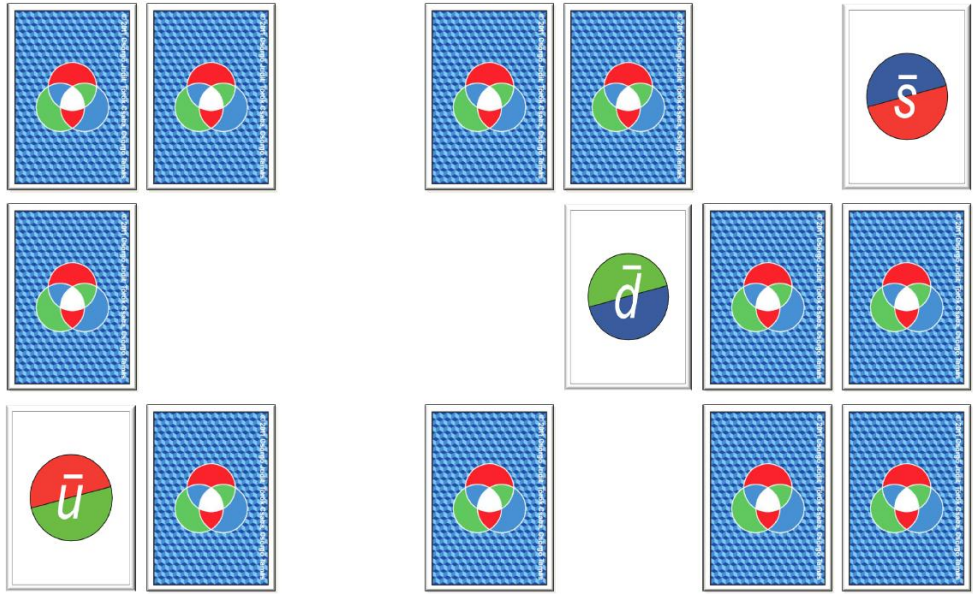

**Figure 5.** Finding a valid antibaryon (a combination of three colored antiquark cards, including each of a red/green, green/blue and blue/red anti-colors), after the baryon of the previous Figure was removed from the middle (or table).

**How to find a Higgs Boson – using Quark Matter Card Game?**

One of the highlights of 2012 in physics is related to two papers, published in August 2012 by the ATLAS and the CMS Collaborations at the CERN LHC accelerator. These two experiments announced, in harmony with each other, the observation of at least one new particle in p+p collisions at CERN LHC [3,4]. It was also shown that at least one of the properties of this new particle is similar to that of the so called Higgs boson, the last and most elusive particle from the Standard Model of particle physics, whose properties are currently under intensive scrutiny in the ATLAS and CMS experiments. This new particle was experimentally found somewhat indirectly, through its decays to directly observable, final state particles. Many of these final state particles are represented in the deck of cards, that represent elementary particles in the Quark Matter Card Game. These Higgs decay properties can now be utilized, playfully, also to include a Higgs boson search to make this Memory style game more exciting. The ATLAS and CMS experiments found a significant signal for at least one new particle, that has at least one property that is similar to the Higgs ($H^0$) boson in the following three different decay modes: $H^0 \to \gamma\gamma$, $H \to Z^0 Z^0 \to l^+ l^- \, l^+ l^-$   $H^0 \to W^+ W^- \to l^+ \nu \, l^- \nu$ . In this notation, charged leptons are summarized as $l^+ = e^+$ or $\mu^+$, $l^- = e^-$ or $\mu^-$, while in the $W^+ W^-$ decay channel, the neutrinos are also those (anti)neutrinos that correspond to the same family as their charged lepton: $\nu = \nu_e$ or $\nu_\mu$ . In the Higgs Boson – On Your Own card game, photons ($\gamma$-s) can also be represented as pairs of charged leptons; thus the $H^0 \to \gamma\gamma$, $H \to Z^0 Z^0$ decays are represented similarly. In addition to the above mentioned the compact notation, let us detail all possible Higgs-boson decays as they may be represented by Quark Matter Cards, as listed in Table 1. This way we can extend and generalize the Memory of Quark Matter

Card Game in a novel way, where the most exciting part of a given play is the search for a combination of four black-and-white cards, that represent a possible Higgs boson decay, corresponding to any of the combinations that are given in the right column of Table 1.

| $H^0$ decay mode | Final state particles/cards |
|---|---|
| $H^0 \to \gamma,\gamma$ or $Z^0 Z^0 \to$ | $e^+e^-$  $e^+e^-$ |
| $\to$ | $e^+e^-$  $\mu^+\mu^-$ |
| $\to$ | $\mu^+\mu^-$  $\mu^+\mu^-$ |
| $H^0 \to W^+W^- \to$ | $e^+\nu_e$  $e^-\bar{\nu}_e$ |
| $\to$ | $e^+\nu_e$  $\mu^-\bar{\nu}_\mu$ |
| $\to$ | $\mu^+\nu_\mu$  $e^-\bar{\nu}_e$ |
| $\to$ | $\mu^+\nu_\mu$  $\mu^-\bar{\nu}_\mu$ |

**Table 1.** Possible Higgs decays as represented in the Higgs Boson – on Your Own Card Game. After turning up two black-and-white cards that can be complemented to any of the four particle cards in the right column of this table by turning up two more cards, the player can "detect a Higgs boson" and win the game, if indeed the all the four particle cards in any line of the right column of this table are found.

This can be achieved by taking some risk: if the first two cards, that were turned face up by a given player can be supplemented by two other cards to form a decay of a Higgs boson (as indicated by the right column of Table 1), however taking this risk may also lead to winning the game on the spot, this way ending a given game without taking all the cards from the middle. If however this attempt to find a Higgs was unsuccessful, the player has to return all the four cards to their original place, in a face down position, and the next player continues to play. It may happen, that in a given game, nobody was able to find a combination corresponding to a Higgs decay. In this case, the game ends by removing all the cards from the middle, combining them to mesons, (anti)baryons and lepton pairs. A small mathematical theorem guarantees that if the game is played with a correct deck of Quark Matter Card Game – Elementary Particles on Your Own [1,2] then by the end of the game, all cards can be picked up, regardless the actual order of picking these cards in a given game. If all cards are used up and the game ended this way, the winner is that person, who collected the largest number of cards, similarly to the situation in the Memory of Quark Matter Card Game.

On a more advanced level, after finding any of the combinations in the right column of Table 1, the player has also to throw a dice and if the result is a "6" than the cards really represent a valid Higgs boson, and the player wins the game, while in all the other cases, when the dice result is 1, 2, 3, 4 or 5, the player found just a so called "background" event. As this is also a valid combination of the cards, the player can keep the cards, but as this is not a real Higgs but just a background event, the given player cannot claim winning the game and has to stop trying:  the next player follows. This way, a realistic element of some luck is introduced to complement the already introduced elements of knowledge and memory.

Two possibilities for a winning, Higgs-decay like configurations are illustrated on Figures 6 and 7, corresponding to lines 2 and 6 of Table 1.

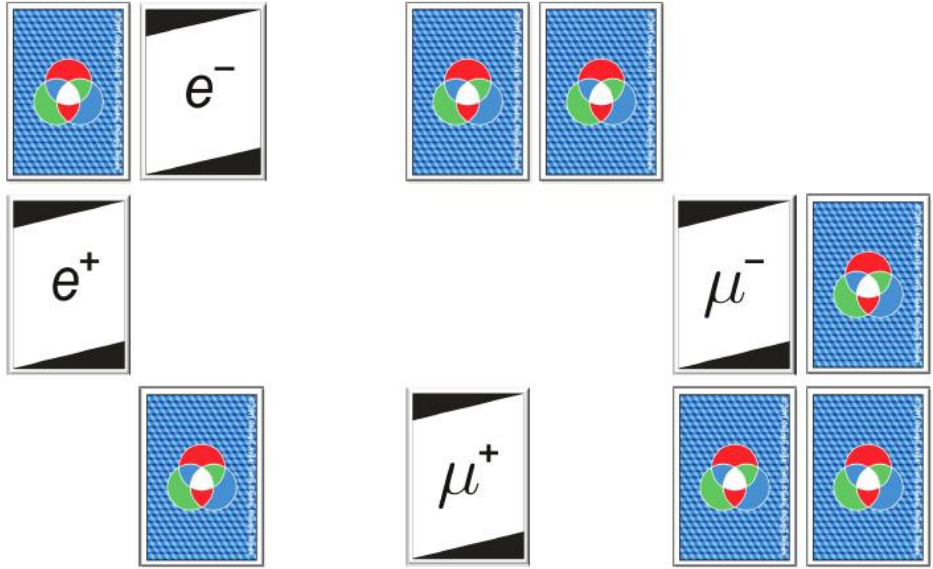

**Figure 6:** After removing the anti-baryon of Figure 5, the player finds a lepton-pair, for example, e+e-, and takes the risk and turns two more cards face up. If these two additional cards represent another lepton pair, (for example, a $\mu^+\mu^-$ pair as on this Figure), then the player found a Higgs boson and won the game.. This process or card combination may represent the final state of both the $H^0 \to \gamma\gamma$ and the $H^0 \to Z^0Z^0$ decays, corresponding to line 2 of Table 1.

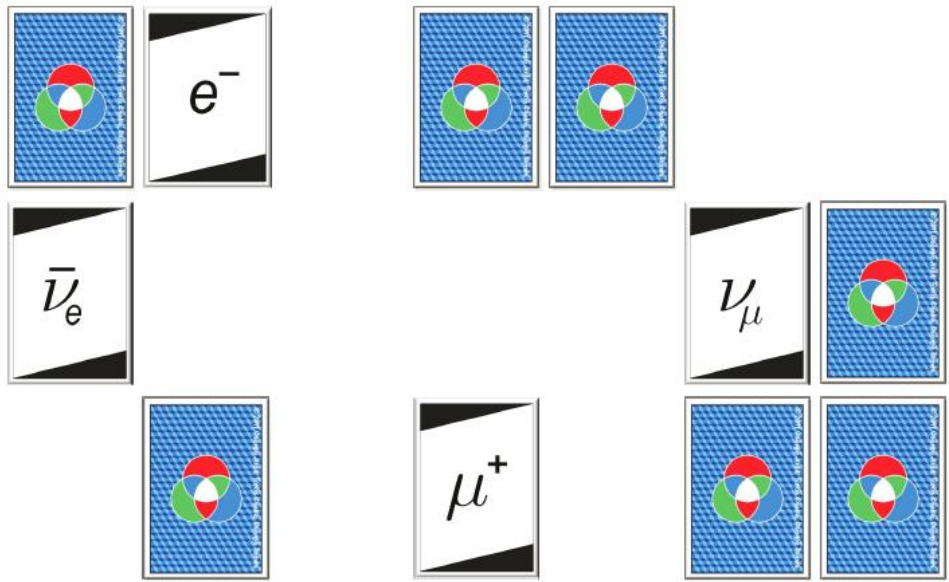

**Figure 7:** After removing the anti-baryon of Figure 5, the player finds a lepton-antilepton combination, for example, a pair of e-$\bar{\nu}_e$, and turns two more cards face up. If these two additional cards also represent leptons in such a way, that one of the decay lines of Table 1 can be completed (for example, by $\mu^+ \nu_\mu$ pair as on this Figure, corresponding to line 6 of Table 1), then the player finds a Higgs boson decay, and wins the game. This process or card combination may represent a final state of a $H^0 \to W^+W^-$ decay.

**Thanks and credits are due** Angela Melocoton of Guests, Users and Visitors (GUV) Center, Brookhaven National Laboratory to for proposing a "Memory" style variation of the Quark Matter card game. This game was further developed and generalized here to the Higgs Boson – on Your Own card game.

**The "Higgs Boson – on Your Own" card game can be played on various levels, similarly to that of the Quark Matter Card Game. Here, a layperson's version of the game was presented.** Three more games, entitled `**ANTI!**', `**Cosmic Showers**' and `**Let's Detect!**' as well as some inspirational physics and the advanced versions of **Quark Matter Card Game** are described in the book entitled:

**Quark Matter Card Game – Elementary Particles on Your Own**,
http://www.lulu.com/spotlight/Reszecskeskartya
© Judit Csörgő, Csaba Török and Tamás Csörgő, 2011

Higgs Boson – on Your Own.
http://www.kfki.hu/~csorgo/talks/12/2012-12-04-TCs-Higgs-boson-on-your-own.pdf
© Tamás Csörgő, 2012